\documentclass[prl,twocolumn,floatfix]{revtex4}

\usepackage{graphicx}
\usepackage{bm}
\usepackage{amsmath,amssymb, amsthm}
\newcommand{\be}{\begin{equation}}
\newcommand{\ee}{\end{equation}}
\newcommand{\ba}{\begin{eqnarray}}
\newcommand{\ea}{\end{eqnarray}}
\newcommand{\nn}{\nonumber \\}
\def\ket#1{\left\vert #1 \right\rangle}

\begin{document}

 \title{Effect of Local Minima on Adiabatic Quantum Optimization}
 \author{M.~H.~S.~Amin}
 \affiliation{D-Wave Systems Inc., 100-4401 Still Creek Drive, Burnaby,
 B.C., V5C 6G9, Canada}

\begin{abstract}

We present a perturbative method to estimate the spectral gap for
adiabatic quantum optimization, based on the structure of the energy
levels in the problem Hamiltonian. We show that for problems that
have exponentially large number of local minima close to the global
minimum, the gap becomes exponentially small making the computation
time exponentially long. The quantum advantage of adiabatic quantum
computation may then be accessed only via the local adiabatic
evolution, which requires phase coherence throughout the evolution
and knowledge of the spectrum. Such problems, therefore, are not
suitable for adiabatic quantum computation.

\end{abstract}

\maketitle

It is widely believed that quantum mechanics can provide speedup for
certain computations. Different quantum algorithms have been
proposed that potentially can solve problems such as factorization
\cite{bib:Shor}, unstructured search \cite{Grover}, or molecular
simulations \cite{Peter} on a quantum computer. One type of problems
for which quantum mechanics may provide an advantage over classical
computation is optimization. In optimization problems, one is
interested in finding solutions that optimize some function subject
to some constraints. Usually, not only the best solution, but also
solutions close to it are of interest.

Physical systems at low temperatures naturally relax to their lowest
energy states, effectively providing optimal solutions to their
energy function. Such a relaxation process, however, may take a very
long time. The time to reach the low energy states may be
significantly reduced via an annealing process in which the
temperature is reduced from a large value to a small value so slowly
that the system stays effectively in equilibrium at all times. The
slow evolution from a thermally disordered to ordered state with
decreasing $T$ will settle the system in one of its low lying energy
states depending on the evolution time. Similar ideas have been
employed in simulated annealing algorithms.

Quantum annealing (QA) \cite{Brooke99,Santoro02} is the quantum
analog of the above classical annealing. In QA the disorder is
introduced quantum mechanically via a Hamiltonian that does not
commute with the optimization Hamiltonian. The added term generally
has a ground state that is a superposition of all the eigenstates of
the optimization Hamiltonian. Therefore, the disordered state is a
superposition rather than a thermal mixture as it is in classical
annealing. The disorder is removed by slowly removing the added term
to the Hamiltonian. The system will then settle into one of its low
lying energy states if the evolution time is long enough.

Closely related to QA is adiabatic quantum computation
(AQC)\cite{Farhi}. In AQC an initial Hamiltonian ${\cal H}_B$ is
slowly deformed into a final (problem) Hamiltonian ${\cal H}_P$:
\ba
 {\cal H} = [1-s(t)]{\cal H}_B + s(t){\cal H}_P, \label{HS}
\ea
with $s(t)$ changing from 0 to 1 between the initial ($t_i{=}0$) and
final ($t_f$) times. In this case, ${\cal H}_B$ plays the role of
the disordering Hamiltonian. The main difference between QA and AQC
is that in the latter, the system is constrained to its ground state
at all times, starting from the ground state of ${\cal H}_B$, into
which it is designed to be initialized, and ending in the ground
state of ${\cal H}_P$, which encodes the solution to the problem of
interest. In other words, AQC is an exact algorithm while QA is
heuristic.

Unlike QA, AQC is not restricted to optimization problems, i.e., the
problem Hamiltonian can be non-diagonal. For example, a universal
AQC can run any quantum algorithm, and has been shown to be
computationally equivalent to the gate model of quantum computation,
as both can be efficiently mapped into each other
\cite{aharonov2004,Lidar06}.

The performance of AQC is determined by the minimum gap $g_m$
between the first two energy levels. In the {\em global} adiabatic
evolution scheme, $s$ is changed uniformly with time ($\dot s =
$const.) and the computation time depends on $g_m$ as $\tau_{\rm
global} \propto g_m^{-2}$. In the local adiabatic scheme
\cite{Roland}, on the other hand, $s$ is a nonlinear function of
time designed in such a way to optimize the computation time by
spending the majority of the evolution time in the vicinity of the
anticrossing. As a result, the computation time of the local AQC is
reduced to $\tau_{\rm local} \propto g_m^{-1}$, which scales better
with $g_m$.

The global and local schemes of AQC are also different in terms of
their response to decoherence. The global scheme is robust against
environmental noise and decoherence
\cite{Childs,Roland2,TAQC,Andy,AA07}. The local adiabatic scheme, on
the other hand, is very sensitive to decoherence. It was shown in
Ref.~\cite{AA07} that in order for the local scheme to change the
scaling of the computation time from $\propto g_m^{-2}$ to  $\propto
g_m^{-1}$, the computation time should be smaller than the global
dephasing time. Moreover, local adiabatic evolution requires
knowledge of the spectrum which is not feasible for general
Hamiltonians.

An important question now is what kind of problems can benefit from
the quantum advantage of AQC without requiring local adiabatic
evolution and therefore phase coherence? It is known that for the
unstructured search problem \cite{Roland}, $\tau_{\rm global} =
O(N)$, which is the complexity of classical search, while $\tau_{\rm
local} = O(\sqrt{N})$, which is the optimal performance of a quantum
algorithm. (Here $N=2^n$, where $n$ is the number of qubits.) Thus,
the advantage over classical computation is only possible via the
local adiabatic evolution. On the other hand, the universal AQC
\cite{aharonov2004,Lidar06,Jake} can provide solution to a problem
in polynomial time if the same problem can be solved in polynomial
time using gate model quantum computation. Evidently, the polynomial
advantage does not depend on local evolution, which can only provide
quadratic enhancement. There have also been previous works to
determine the complexity of AQC for some other special Hamiltonians
\cite{Farhi05,Vazirani,Mosca07}. In this letter, we study this
problem for a rather general form of adiabatic quantum optimization.

We consider physically realizable initial and final Hamiltonians:
 \ba
 {\cal H}_B &=& -{\Delta\over 2}\sum_{i=1}^n \sigma^x_i, \label{HB} \\
 {\cal H}_P &=& -{{\cal E}\over 2} \left[ \sum_{i=1}^n h_i\sigma^z_i
 + \sum_{i,j=1}^n J_{ij} \sigma^z_i\sigma^z_j \right], \label{HP}
 \ea
where $\sigma^{x,z}_i$ are the Pauli matrices for the $i$-th qubit,
$h_i$ and $J_{ij}$ are dimensionless local fields and coupling
coefficients respectively [typically $O(1)$], and ${\cal E}$ is some
characteristic energy scale for ${\cal H}_P$. The initial
Hamiltonian ${\cal H}_B$ has a nondegenerate ground state
$|\psi_G(0)\rangle=|\overline{0^n}\rangle$. Here, we have adopted
the notation $|\bar{z}\rangle = H^{\otimes n}|z\rangle$, $z\in
\{0,1\}^n$, for states that are diagonal in the Hadamard basis, with
$H$ being the Hadamard transformation.

We denote the ground state of the total Hamiltonian by
$|\psi_G\rangle = \sum_z a_z |z\rangle$, where $a_z$ are complex
probability amplitudes. At the beginning of the evolution,
$a_z=1/\sqrt{N}$, therefore $|\psi_G\rangle$ is a uniform
superposition of all the states in the computation basis, but at the
end of the evolution, it is only a superposition of the final
solutions. The transition from large to small superpositions happens
very suddenly at the minimum gap, which in the limit $g_m\to 0$
represents a quantum phase transition. Here, we only focus on
first-order phase transition in which the gap is in the form of an
avoided crossing \cite{Schaller07}. If $g_m$ is much smaller than
the separation of the two crossing levels from other energy levels,
then the slow evolution of the system close to the anticrossing will
be restricted only to those levels. Using a new coordinate $\epsilon
= 2E(s{-}s^*)$, where $E$ is an energy scale characterizing the
anticrossing and $s^*$ is its position, one can write a two-state
Hamiltonian:
\be {\cal H} =-(\epsilon \tau_z + g_m \tau_x )/2,
 \label{H2L}
\ee
%
with $\tau_{x,z}$ being the Pauli matrices in the two-state
subspace.

Immediately before and after the anticrossing, we write
 \be
 |\psi_G^\pm \rangle = |\psi_G(\pm \epsilon_0) \rangle =
 \sum_{z\in \{0,1\}^n} a_z^\pm |z\rangle,
 \ee
with $\epsilon_0 \ll E_{12}$, and $E_{12}$ being the energy
separation between the first two excited states. Using (\ref{H2L})
it is easy to show that for $\epsilon_0 \gg g_m$
 \be
 g_m \approx \epsilon_0 |\langle\psi_G^+|\psi_G^- \rangle|,
 \label{gm0}
 \ee
i.e., $g_m$ is proportional to the overlap of the wave-functions
before and after the anticrossing.

Let us introduce two sets
 \be
 S^\pm = \{ z :\ |a_z^\pm| > \delta \}, \label{Spm}
 \ee
where $\delta$ is a small number. Since all elements in $S^\pm$
contribute to the superposition, the normalization condition
requires $|a_z^\pm|=O(1/\sqrt{|S^\pm|})$, yielding
 \be
 |\psi_G^\pm \rangle \sim
 \sum_{z\in S^\pm}^N {1\over\sqrt{|S^\pm|}} |z\rangle,
 \ee
where $|S|$ denotes the cardinality of set $S$. The minimum gap will
therefore be
 \be
 g_m \propto {|S^+ {\cap} S^-| \over\sqrt{|S^+||S^-|}} \leq \sqrt{|S^+|\over |S^-|}.
 \label{gmin}
 \ee
The equality happens when all states in $S^+$ also belong to $S^-$.
We shall only focus on this case as it provides an upper limit for
$g_m$. In order to understand what can make the gap small, we need
to understand how $S^\pm$ are constructed. To this end, we use
perturbation expansion.

We introduce the dimensionless parameter
 \be
 \zeta(t) \equiv {s(t){\cal E} \over [1-s(t)]\Delta},
 \ee
which varies from 0 to $\infty$ during the evolution. We shall drop
the time dependence of $\zeta$ for simplicity. We start by
calculating $|\psi_G\rangle$ near the end of the evolution, where
$\zeta$ is large, by considering ${\cal H}_{1/\zeta} = {\cal H}_P +
(1/\zeta) {\cal H}_B$. Let us for now assume that the problem has a
unique solution, therefore the ground state of ${\cal H}_P$ is a
nondegenerate state $\ket{f}$. To the 0-th order in $1/\zeta$,
$|\psi_G^{(0)}\rangle = |f\rangle$. Since ${\cal H}_B$ is a linear
function of $\sigma_i^x$, it can only generate single qubit flips.
Thus, to include a state $\ket{z}$ with Hamming distance $m=||z-f||$
from the solution $\ket{f}$, in $|\psi_G\rangle$, we need to apply
${\cal H}_B$ at least $m$ times. Therefore, $a_z^+$ is nonzero only
after the $m$-th order perturbation: $a_z^+ {=} O(1/\zeta^m)$. This
restricts the states in $S^+$ to be close to $f$ in Hamming
distance. Requiring $1/\zeta^m {\gtrsim} \delta$, we find
 \ba
 S^+ &\approx& \{ z : \ ||z-f|| < m_c \}, \\
 m_c &\propto& {\log(1/\delta)\over \log \zeta^+},
 \qquad \zeta^\pm \equiv \zeta(\epsilon{=}\pm \epsilon_0). \label{mc}
 \ea

The above argument can be easily generalized to multi-solution
problems by writing the unperturbed ground state near the end of the
evolution as $|\psi_G^{(0)}\rangle {=} N_s^{-1/2}
\sum_{l=1}^{N_s}|f_l\rangle$, where $f_l$ is the $l$-th solution
among the total $N_s$ solutions of the problem. In this case,
 \ba
 S^+ &\approx& \{ z : \ \min_l ||z-f_l|| < m_c \}. \label{Sp}
 \ea
Therefore, {\em the set $S^+$ is constructed from states that are
close in Hamming distance to the final solutions}.

To find $S^-$, we perform perturbation expansion around ${\cal
H}_B$, with $\zeta$ as the small parameter, using ${\cal H}_{\zeta}
= {\cal H}_B + \zeta {\cal H}_P$. Before performing the perturbation
expansion, let us use our intuition to understand how $S^-$ can be
formed. At $\zeta=0$, the ground state of the Hamiltonian ${\cal
H}_B$ is the uniform superposition of all the states in the
computation basis. Adding a small perturbation $\zeta {\cal H}_P$ to
the Hamiltonian will introduce a penalty to those eigenstates of
${\cal H}_P$ that have large eigenvalues. As a result, one expects
that adding $\zeta {\cal H}_P$ will remove those high energy
eigenstates from the superposition. The larger the $\zeta$, the more
high energy levels will be removed from the superposition and
eventually only low lying states will survive.

Let us make this intuitive argument more quantitative. Since ${\cal
H}_B$ is diagonal in the Hadamard basis, we need to do the
perturbation expansion in that basis. Let us write
 \be
 |\psi_G\rangle = \sum_{z\in \{0,1\}^n} b_{\bar{z}} |\bar{z}\rangle.
 \ee
To the 0-th order, the wave function is $|\psi_G^{(0)}\rangle =
|\overline{0^n}\rangle$. The Hamiltonian ${\cal H}_P$ is a bilinear
function of $\sigma_i^z$, hence it can generate single and two qubit
flips in the Hadamard basis. Again, in order to include a state
$\ket{\bar{z}}$ into the superposition $|\psi_G\rangle$, where $z$
has a Hamming weight $w$, we need to apply ${\cal H}_B$ at least
$w/2$ times. This requires $b_{\bar{z}} = O(\zeta^{w/2})$ thereby
making $b_{\bar{z}}$ non-negligible only if
 \be
 w < w_c \propto {\log(1/\delta) \over \log(1/\zeta^-)}.
 \ee

In order to determine $S^-$, we need to know how $|\psi_G^- \rangle$
is formed in the computation basis, not in the Hadamard basis. Since
${\cal H}_B$ does not commute with ${\cal H}_P$, one can use the
uncertainty principle to find the restriction imposed by the
perturbation in the Hadamard basis, on the wave function in the
computation basis. Let $\delta E_P$ and $\delta E_B$ represent
uncertainties in ${\cal H}_P$ and ${\cal H}_B$, respectively:
 \ba
 &&\delta E_B \equiv (\langle {\cal H}_B^2 \rangle - \langle {\cal H}_B
 \rangle^2)^{1/2}, \nn
 &&\delta E_P \equiv (\langle {\cal H}_P^2 \rangle - \langle {\cal H}_P
 \rangle^2)^{1/2},
 \ea
where $\langle ... \rangle$ represents expectation value. The
uncertainty principle requires $\delta E_B \cdot \delta E_P \geq
{1\over 2} \langle i[{\cal H}_B,{\cal H}_P] \rangle$. Every state
$\ket{\bar{z}}$ is an eigenstate of ${\cal H}_B$ with eigenvalue
$w\Delta$, where $w$ is the Hamming weight of $z$.  For the ground
state $|\psi_G\rangle$, therefore, we have $\delta E_B \sim w_c
\Delta$. The uncertainty principle requires $\delta E_P \propto
1/w_c$, leading to
 \ba
 S^- &\approx& \{ z: \ E_z < E_c \}, \label{Sm} \\
 E_c &\propto& {\log(1/\zeta^-)\over \log(1/\delta) }. \label{Ec}
 \ea
As expected, {\em the set $S^-$ is made of low energy eigenstates of
${\cal H}_P$}.

Equations (\ref{mc}) and (\ref{Ec}) suggest that as $\zeta^\pm \to
1$, $m_c \to \infty$ and $E_c \to 0$. Since the perturbation
expansion breaks down at $\zeta \sim 1$, these equations cannot be
extended all the way to $\zeta \sim 1$. In fact, $\zeta \sim 1$ is
exactly where the quantum phase transition and therefore the
anticrossing occurs. However, to calculate $g_m$ using (\ref{gm0}),
we need $\epsilon_0 \gg g_m$, which ensures that
$|\psi_G^\pm\rangle$ are indeed defined far away from the phase
transition point, where the perturbation expansion and therefore
(\ref{Sp}) and (\ref{Sm}) still hold. The important fact to notice
is that the sets $S^\pm$ are formed in completely different ways:
$S^-$ is constructed by all the energy levels below some threshold,
while $S^+$ is formed by all the states in Hamming proximity to the
answers. The two sets could be very different leading to a very
small energy gap.

For the upper limit in (\ref{gmin}), i.e., $g_m \propto
\sqrt{|S^+|/|S^-|}$, the computation time will be $\tau_{\rm local}
\propto \sqrt{\tau_{\rm global}} \propto \sqrt{|S^-|/|S^+|}$. If the
problem Hamiltonian happens to have an exponentially large number of
low energy states that have large Hamming distances to the correct
solutions (i.e., low energy {\em local minima}), those states will
belong to $S^-$ and not to $S^+$. The resulting gap will therefore
be exponentially small, and the computation time will be extremely
large. Especially, if $|S^-|$ becomes a fraction of $N$, then
$\tau_{\rm global} = O(N)$, which is the complexity of the
exhaustive search. The quantum advantage then will only be
achievable via the local adiabatic scheme for which $\tau_{\rm
local} = O(\sqrt{N})$.

An interesting example of such difficult problems is random
3-Satisfiability problem (3-SAT). These problems exhibit a phase
transition when the ratio of the number of clauses $m$ to the number
of variables $n$ reaches $\approx 4.2$ \cite{3SATphtrans}. Before
the phase transition the number of solutions that satisfy the 3-SAT
formula is extremely large, but suddenly after the phase transition
point the number of satisfying solutions drops to zero. Therefore,
near the phase transition point, by adding a few clauses to the
3-SAT formula, a large number of states that did satisfy it before
will no longer do so. Those solutions, however, violate only those
few clauses. In terms of energy, by adding a few terms in the
Hamiltonian that provide penalties for those few clauses, an
exponentially large number of states that used to be global minima
suddenly become local minima but with energies very close to the
ground state energy. This, therefore, would result in an
exponentially small gap, as confirmed numerically
\cite{Znidaric2005} and analytically \cite{Znidaric2006}.

Another example is spin glasses \cite{Fischer}, in which the local
fields $h_i$ are small or zero and coupling coefficients $J_{ij}$
randomly couple (only) neighboring qubits. In such problems, domains
can be formed if a large number of physically close qubits are
strongly coupled to each other \cite{Kaminsky07}. The qubits within
a domain minimize the coupling terms $J_{ij}$ in ${\cal H}_P$. Those
terms, however, remain unchanged if all of the qubits in the domain
flip together. If the energy cost of flipping a domain, imposed by
the field terms ($h_i$) and by the violation of the bounds
($J_{ij}$) at the domain boundary is not so large, then such a
domain flipped state will form a low energy local minimum. If the
size of the domain is large, then the Hamming distance between the
local minimum and the global one will also be large. Thus, the local
minimum and all the states close to it do not belong to $S^+$, while
they do belong to $S^-$. A large number of such domains may make
$g_m$ exponentially small. It should be mentioned that if $h_i=0$,
then the final Hamiltonian will be symmetric under the total spin
flip operation and the phase transition is likely to be second
order, invalidating our assumption.

To conclude, we have used a perturbative approach to estimate the
gap size for adiabatic quantum optimization problems. The gap is
found to be inversely proportional to the square root of the number
of states that have energies close to the global minimum. Therefore,
problems that have a large number of low energy local minima tend to
have a small gap. In general, problem instances in which the
interaction terms in the Hamiltonian dominate the energy eigenvalues
(i.e., typical values of $J_{ij}$ are much larger than those for
$h_i$) are likely to form low energy local minima and therefore make
the gap small. If the number of low energy local minima becomes
exponentially large, then the gap will be exponentially small. In
such cases, only a local adiabatic evolution scheme can provide
quantum advantage over classical computation. Local AQC, however,
requires phase coherence during the evolution \cite{AA07} and
knowledge of the energy spectrum which limits its practicality.
These problems, although unsuitable for AQC, could be suitable for
heuristic algorithms (if approximate solutions are acceptable),
because the chance of finding a solution within the acceptance
threshold will be large. Quantum annealing therefore may provide
good enough solutions in a short time, although finding the global
minimum via AQC can take an extremely long time.

Finally, it should be mentioned that the energy gaps considered here
are only the avoided crossing type, which correspond to first-order
quantum phase transitions. If the final Hamiltonian possesses a
symmetry that imposes a spontaneous symmetry breaking at the
anticrossing, the resulting phase transition may become second order
or higher orders. It has been stated that Hamiltonians with higher
order phase transitions can provide better (possibly polynomial)
scaling with the number of qubits \cite{Schutzhold06,Schaller07}.
Study of those is beyond the scope of the present paper.

The author is grateful to A.J.~Berkley, V.~Choi, F.~Chudak,
R.~Harris, W.M.~Kaminsky, W.G.~Macready, G.~Rose, and C.J.S. Truncik
for discussions.

\bibliography{LMinima}

\begin{thebibliography}{25}
\expandafter\ifx\csname natexlab\endcsname\relax\def\natexlab#1{#1}\fi
\expandafter\ifx\csname bibnamefont\endcsname\relax
  \def\bibnamefont#1{#1}\fi
\expandafter\ifx\csname bibfnamefont\endcsname\relax
  \def\bibfnamefont#1{#1}\fi
\expandafter\ifx\csname citenamefont\endcsname\relax
  \def\citenamefont#1{#1}\fi
\expandafter\ifx\csname url\endcsname\relax
  \def\url#1{\texttt{#1}}\fi
\expandafter\ifx\csname urlprefix\endcsname\relax\def\urlprefix{URL }\fi
\providecommand{\bibinfo}[2]{#2}
\providecommand{\eprint}[2][]{\url{#2}}

\bibitem[{\citenamefont{Shor}(1997)}]{bib:Shor}
\bibinfo{author}{\bibfnamefont{P.}~\bibnamefont{Shor}}, \bibinfo{journal}{SIAM
  J. Comput.} \textbf{\bibinfo{volume}{26}}, \bibinfo{pages}{1484}
  (\bibinfo{year}{1997}).

\bibitem[{\citenamefont{Grover}(1997)}]{Grover}
\bibinfo{author}{\bibfnamefont{L.~K.} \bibnamefont{Grover}},
  \bibinfo{journal}{Phys. Rev. Lett.} \textbf{\bibinfo{volume}{79}},
  \bibinfo{pages}{325} (\bibinfo{year}{1997}).

\bibitem[{\citenamefont{{Aspuru-Guzik}
  et~al.}(2005)\citenamefont{{Aspuru-Guzik}, {Dutoi}, {Love}, and
  {Head-Gordon}}}]{Peter}
\bibinfo{author}{\bibfnamefont{A.}~\bibnamefont{{Aspuru-Guzik}}},
  \bibinfo{author}{\bibfnamefont{A.~D.} \bibnamefont{{Dutoi}}},
  \bibinfo{author}{\bibfnamefont{P.~J.} \bibnamefont{{Love}}},
  \bibnamefont{and}
  \bibinfo{author}{\bibfnamefont{M.}~\bibnamefont{{Head-Gordon}}},
  \bibinfo{journal}{Science} \textbf{\bibinfo{volume}{309}},
  \bibinfo{pages}{1704} (\bibinfo{year}{2005}).

\bibitem[{\citenamefont{{Brooke} et~al.}(1999)\citenamefont{{Brooke}, {Bitko},
  {Rosenbaum}, and {Aeppli}}}]{Brooke99}
\bibinfo{author}{\bibfnamefont{J.}~\bibnamefont{{Brooke}}},
  \bibinfo{author}{\bibfnamefont{D.}~\bibnamefont{{Bitko}}},
  \bibinfo{author}{\bibfnamefont{T.~F.} \bibnamefont{{Rosenbaum}}},
  \bibnamefont{and} \bibinfo{author}{\bibfnamefont{G.}~\bibnamefont{{Aeppli}}},
  \bibinfo{journal}{Science} \textbf{\bibinfo{volume}{284}},
  \bibinfo{pages}{779} (\bibinfo{year}{1999}).

\bibitem[{\citenamefont{{Santoro} et~al.}(2002)\citenamefont{{Santoro},
  {Marto{\v n}{\'a}k}, {Tosatti}, and {Car}}}]{Santoro02}
\bibinfo{author}{\bibfnamefont{G.~E.} \bibnamefont{{Santoro}}},
  \bibinfo{author}{\bibfnamefont{R.}~\bibnamefont{{Marto{\v n}{\'a}k}}},
  \bibinfo{author}{\bibfnamefont{E.}~\bibnamefont{{Tosatti}}},
  \bibnamefont{and} \bibinfo{author}{\bibfnamefont{R.}~\bibnamefont{{Car}}},
  \bibinfo{journal}{Science} \textbf{\bibinfo{volume}{295}},
  \bibinfo{pages}{2427} (\bibinfo{year}{2002}).

\bibitem[{\citenamefont{Farhi et~al.}(2001)\citenamefont{Farhi, Goldstone,
  Gutmann, Lapan, Lundgren, and Preda}}]{Farhi}
\bibinfo{author}{\bibfnamefont{E.}~\bibnamefont{Farhi}},
  \bibinfo{author}{\bibfnamefont{J.}~\bibnamefont{Goldstone}},
  \bibinfo{author}{\bibfnamefont{S.}~\bibnamefont{Gutmann}},
  \bibinfo{author}{\bibfnamefont{J.}~\bibnamefont{Lapan}},
  \bibinfo{author}{\bibfnamefont{A.}~\bibnamefont{Lundgren}}, \bibnamefont{and}
  \bibinfo{author}{\bibfnamefont{D.}~\bibnamefont{Preda}},
  \bibinfo{journal}{Science} \textbf{\bibinfo{volume}{292}},
  \bibinfo{pages}{472} (\bibinfo{year}{2001}).

\bibitem[{\citenamefont{Aharonov et~al.}(2007)\citenamefont{Aharonov, van Dam,
  Kempe, Landau, and Lloyd}}]{aharonov2004}
\bibinfo{author}{\bibfnamefont{D.}~\bibnamefont{Aharonov}},
  \bibinfo{author}{\bibfnamefont{W.}~\bibnamefont{van Dam}},
  \bibinfo{author}{\bibfnamefont{J.}~\bibnamefont{Kempe}},
  \bibinfo{author}{\bibfnamefont{Z.}~\bibnamefont{Landau}}, \bibnamefont{and}
  \bibinfo{author}{\bibfnamefont{S.}~\bibnamefont{Lloyd}},
  \bibinfo{journal}{SIAM J. Comput.} \textbf{\bibinfo{volume}{37}},
  \bibinfo{pages}{166} (\bibinfo{year}{2007}).

\bibitem[{\citenamefont{{Mizel} et~al.}(2007)\citenamefont{{Mizel}, {Lidar},
  and {Mitchell}}}]{Lidar06}
\bibinfo{author}{\bibfnamefont{A.}~\bibnamefont{{Mizel}}},
  \bibinfo{author}{\bibfnamefont{D.~A.} \bibnamefont{{Lidar}}},
  \bibnamefont{and}
  \bibinfo{author}{\bibfnamefont{M.}~\bibnamefont{{Mitchell}}},
  \bibinfo{journal}{Phys. Rev. Lett.} \textbf{\bibinfo{volume}{99}},
  \bibinfo{pages}{070502} (\bibinfo{year}{2007}).

\bibitem[{\citenamefont{Roland and Cerf}(2002)}]{Roland}
\bibinfo{author}{\bibfnamefont{J.}~\bibnamefont{Roland}} \bibnamefont{and}
  \bibinfo{author}{\bibfnamefont{N.~J.} \bibnamefont{Cerf}},
  \bibinfo{journal}{Phys. Rev. A} \textbf{\bibinfo{volume}{65}},
  \bibinfo{pages}{042308} (\bibinfo{year}{2002}).

\bibitem[{\citenamefont{Childs et~al.}(2001)\citenamefont{Childs, Farhi, and
  Preskill}}]{Childs}
\bibinfo{author}{\bibfnamefont{A.~M.} \bibnamefont{Childs}},
  \bibinfo{author}{\bibfnamefont{E.}~\bibnamefont{Farhi}}, \bibnamefont{and}
  \bibinfo{author}{\bibfnamefont{J.}~\bibnamefont{Preskill}},
  \bibinfo{journal}{Phys. Rev. A} \textbf{\bibinfo{volume}{65}},
  \bibinfo{pages}{012322} (\bibinfo{year}{2001}).

\bibitem[{\citenamefont{Roland and Cerf}(2005)}]{Roland2}
\bibinfo{author}{\bibfnamefont{J.}~\bibnamefont{Roland}} \bibnamefont{and}
  \bibinfo{author}{\bibfnamefont{N.~J.} \bibnamefont{Cerf}},
  \bibinfo{journal}{Phys. Rev. A} \textbf{\bibinfo{volume}{71}},
  \bibinfo{pages}{032330} (\bibinfo{year}{2005}).

\bibitem[{\citenamefont{Amin et~al.}()\citenamefont{Amin, Love, and
  Truncik}}]{TAQC}
\bibinfo{author}{\bibfnamefont{M.~H.~S.} \bibnamefont{Amin}},
  \bibinfo{author}{\bibfnamefont{P.~J.} \bibnamefont{Love}}, \bibnamefont{and}
  \bibinfo{author}{\bibfnamefont{C.~J.~S.} \bibnamefont{Truncik}},
  \eprint{cond-mat/0609332}.

\bibitem[{\citenamefont{Wan et~al.}()\citenamefont{Wan, Amin, and Wang}}]{Andy}
\bibinfo{author}{\bibfnamefont{A.~T.~S.} \bibnamefont{Wan}},
  \bibinfo{author}{\bibfnamefont{M.~H.~S.} \bibnamefont{Amin}},
  \bibnamefont{and} \bibinfo{author}{\bibfnamefont{S.~X.} \bibnamefont{Wang}},
  \eprint{cond-mat/0703085}.

\bibitem[{\citenamefont{Amin and Averin}()}]{AA07}
\bibinfo{author}{\bibfnamefont{M.~H.~S.} \bibnamefont{Amin}} \bibnamefont{and}
  \bibinfo{author}{\bibfnamefont{D.~V.} \bibnamefont{Averin}},
  \bibinfo{note}{arXiv:0708.0384}.

\bibitem[{\citenamefont{Biamonte and Love}()}]{Jake}
\bibinfo{author}{\bibfnamefont{J.~D.} \bibnamefont{Biamonte}} \bibnamefont{and}
  \bibinfo{author}{\bibfnamefont{P.~J.} \bibnamefont{Love}},
  \bibinfo{note}{arXiv:0704.1287}.

\bibitem[{\citenamefont{Farhi et~al.}()\citenamefont{Farhi, Goldstone, Gutmann,
  and Nagaj}}]{Farhi05}
\bibinfo{author}{\bibfnamefont{E.}~\bibnamefont{Farhi}},
  \bibinfo{author}{\bibfnamefont{J.}~\bibnamefont{Goldstone}},
  \bibinfo{author}{\bibfnamefont{S.}~\bibnamefont{Gutmann}}, \bibnamefont{and}
  \bibinfo{author}{\bibfnamefont{D.}~\bibnamefont{Nagaj}},
  \eprint{quant-ph/0512159}.

\bibitem[{\citenamefont{Ioannou and Mosca}()}]{Mosca07}
\bibinfo{author}{\bibfnamefont{L.~M.} \bibnamefont{Ioannou}} \bibnamefont{and}
  \bibinfo{author}{\bibfnamefont{M.}~\bibnamefont{Mosca}},
  \eprint{quant-ph/0702241}.

\bibitem[{\citenamefont{van Dam et~al.}()\citenamefont{van Dam, Mosca, and
  Vazirani}}]{Vazirani}
\bibinfo{author}{\bibfnamefont{W.}~\bibnamefont{van Dam}},
  \bibinfo{author}{\bibfnamefont{M.}~\bibnamefont{Mosca}}, \bibnamefont{and}
  \bibinfo{author}{\bibfnamefont{U.}~\bibnamefont{Vazirani}},
  \eprint{Proceedings of the 42nd Annual Symposium on Foundations of Computer
  Science, pp. 279-287 (2001)}.

\bibitem[{\citenamefont{{Schaller} and {Sch{\"u}tzhold}}()}]{Schaller07}
\bibinfo{author}{\bibfnamefont{G.}~\bibnamefont{{Schaller}}} \bibnamefont{and}
  \bibinfo{author}{\bibfnamefont{R.}~\bibnamefont{{Sch{\"u}tzhold}}},
  \eprint{arXiv:0708.1882}.

\bibitem[{\citenamefont{Kirkpatrick and Selman}(1994)}]{3SATphtrans}
\bibinfo{author}{\bibfnamefont{S.}~\bibnamefont{Kirkpatrick}} \bibnamefont{and}
  \bibinfo{author}{\bibfnamefont{B.}~\bibnamefont{Selman}},
  \bibinfo{journal}{Science} \textbf{\bibinfo{volume}{264}},
  \bibinfo{pages}{1297} (\bibinfo{year}{1994}).

\bibitem[{\citenamefont{\v{Z}nidari\v{c}}(2005)}]{Znidaric2005}
\bibinfo{author}{\bibfnamefont{M.}~\bibnamefont{\v{Z}nidari\v{c}}},
  \bibinfo{journal}{Phys. Rev. A} \textbf{\bibinfo{volume}{71}},
  \bibinfo{pages}{062305} (\bibinfo{year}{2005}).

\bibitem[{\citenamefont{\v{Z}nidari\v{c} and Horvat}(2006)}]{Znidaric2006}
\bibinfo{author}{\bibfnamefont{M.}~\bibnamefont{\v{Z}nidari\v{c}}}
  \bibnamefont{and} \bibinfo{author}{\bibfnamefont{M.}~\bibnamefont{Horvat}},
  \bibinfo{journal}{Phys. Rev. A} \textbf{\bibinfo{volume}{73}},
  \bibinfo{pages}{022329} (\bibinfo{year}{2006}).

\bibitem[{\citenamefont{Fischer and Hertz}(1991)}]{Fischer}
\bibinfo{author}{\bibfnamefont{K.~H.} \bibnamefont{Fischer}} \bibnamefont{and}
  \bibinfo{author}{\bibfnamefont{J.~A.} \bibnamefont{Hertz}},
  \emph{\bibinfo{title}{Spin glasses}} (\bibinfo{publisher}{Cambridge
  University Press}, \bibinfo{year}{1991}).

\bibitem[{\citenamefont{Kaminsky and Lloyd}()}]{Kaminsky07}
\bibinfo{author}{\bibfnamefont{W.~M.} \bibnamefont{Kaminsky}} \bibnamefont{and}
  \bibinfo{author}{\bibfnamefont{S.}~\bibnamefont{Lloyd}}, \eprint{in
  preparation}.

\bibitem[{\citenamefont{{Sch{\"u}tzhold} and {Schaller}}(2006)}]{Schutzhold06}
\bibinfo{author}{\bibfnamefont{R.}~\bibnamefont{{Sch{\"u}tzhold}}}
  \bibnamefont{and}
  \bibinfo{author}{\bibfnamefont{G.}~\bibnamefont{{Schaller}}},
  \bibinfo{journal}{\pra} \textbf{\bibinfo{volume}{74}},
  \bibinfo{pages}{060304} (\bibinfo{year}{2006}).

\end{thebibliography}

\end{document}